\documentclass[aps,prb,twocolumn,superscriptaddress, show pacs]{revtex4-1}
\usepackage{bm, amsmath}
\usepackage{graphicx}
\usepackage{color}
\usepackage{epstopdf}
\usepackage{graphicx}
\usepackage{dcolumn}
\usepackage{bm}
\usepackage{subfigure}

\begin{document}

\title{Three-dimensional  chiral skyrmions with attractive interparticle interactions}


\author{A~O~Leonov}
\thanks{A.Leonov@ifw-dresden.de}
\affiliation{Center for Chiral Science, Hiroshima University, Higashi-Hiroshima, 
Hiroshima 739-8526, Japan}
\affiliation{IFW Dresden, Postfach 270016, D-01171 Dresden, Germany}   
\affiliation{Zernike Institute for Advanced Materials, University of Groningen, 
Groningen, 9700AB, The Netherlands}

\author{T~L~Monchesky}
\affiliation{Center for Chiral Science, Hiroshima University, Higashi-Hiroshima, 
Hiroshima 739-8526, Japan}
\affiliation{Department of Physics and Atmospheric Science, Dalhousie University, 
Halifax, Nova Scotia, Canada B3H 3J5}

\author{J~C~Loudon}
\affiliation{Department of Materials Science and
  Metallurgy, 27 Charles Babbage Road, Cambridge, CB3 0FS, United Kingdom}

\author{A~N~Bogdanov}
\affiliation{Center for Chiral Science, Hiroshima University, Higashi-Hiroshima, 
Hiroshima 739-8526, Japan}
\affiliation{IFW Dresden, Postfach 270016, D-01171 Dresden, Germany}

\date{\today}

\begin{abstract}
{ We introduce a new class of isolated three-dimensional
 skyrmion that can occur within the cone  phase
of chiral magnetic materials.
These novel solitonic states consist of an   axisymmetric 
core separated from the host phase by an  asymmetric
  shell. These skyrmions attract one another. 
We derive regular solutions for isolated  skyrmions arising in the cone
phase of cubic helimagnets and investigate their bound states.
}
\end{abstract}

         
\maketitle

The \textit{Dzyaloshinskii-Moriya} (DM)  interaction 
in noncentrosymmetric magnets  is  a result of their
crystallographic handedness \cite{Dz64} and  is  responsible for
the formation of long-range modulations with a fixed sense of 
the magnetization rotation \cite{Dz64,Bak80} and
the stabilization of two-dimensional axisymmetric
localized structures  called \textit{skyrmions}  
\cite{JETP89,JMMM94}. 
Long-range homochiral modulations (\textit{helices}) were found in
 the  noncentrosymmetric cubic ferromagnet MnSi several
decades ago and  since then, other  cubic ferromagnets with
B20 structures  have been investigated intensively
 \cite{Ishikawa76,Beille81,Lebech89}  as have 
other chiral magnetic materials \cite{Togawa12,Porter15}.
Isolated chiral skyrmions have been discovered recently
in PdFe/Ir(111) nanolayers with induced 
DM interactions and strong easy-axis anisotropy
\cite{Romming13,Romming15,Leonov15}.

Chiral skyrmions are two-dimensional topological solitons with 
an axisymmetric structure localized in nanoscale cylindrical regions. 
They exist as ensembles of weakly
repulsive particles in the \textit{saturated} phase of
noncentrosymmetric magnets \cite{JMMM94,JETPL95,JPCS11}  in
  which all the atomic spins are parallel to an applied magnetic
  field. In cubic helimagnets, below a certain critical field, $H_D$,
the saturated phase transforms into the chiral helical state with the
propagation direction along the applied field  called the 
\textit{cone} phase \cite{Bak80}.  Unlike  the saturated
phase, the boundary values imposed by the longitudinal modulations of
the cone phase violate a rotational symmetry of the system, and are thus 
incompatible with the axisymmetric arrangement of
skyrmions investigated  in Refs. 
\onlinecite{JMMM94,JETPL95,Romming13,Romming15}.  
Then, the question arises: \textit{``are there any  localized  states
  compatible with the encompassing cone phase?'' }

\begin{figure}
\includegraphics[width=0.8\columnwidth]{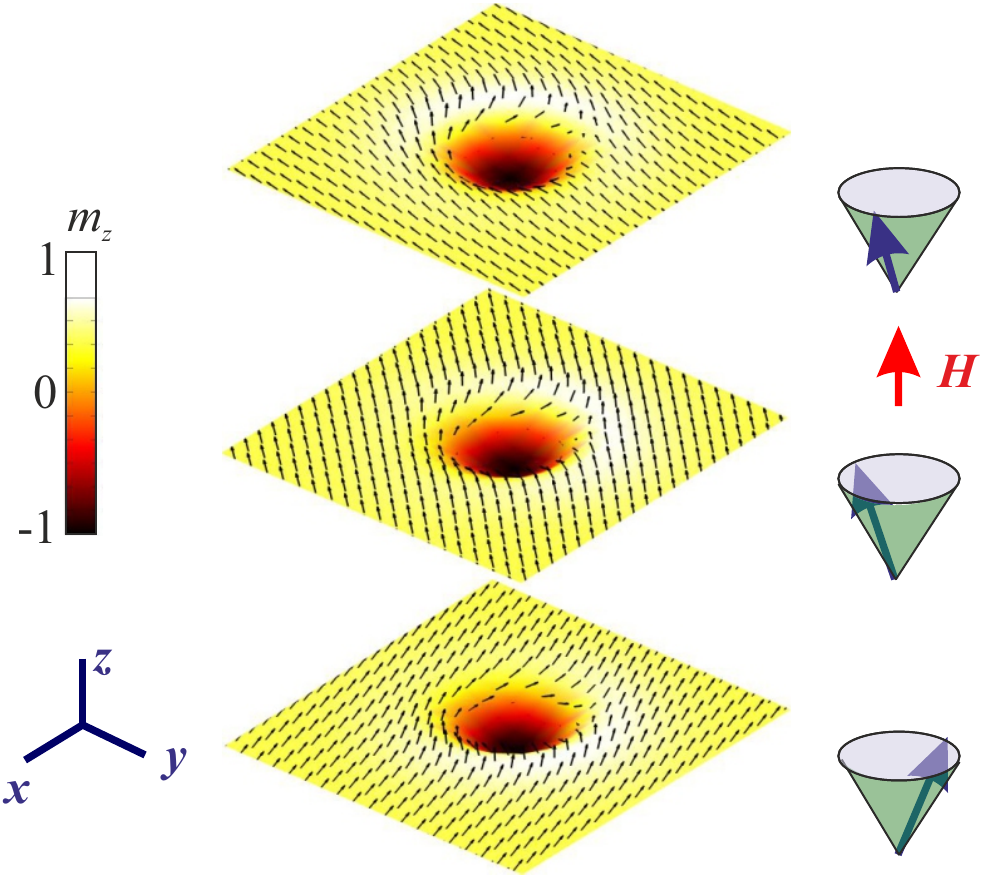}
\caption{ 
(color online). Magnetic structure of an isolated skyrmion in the cone phase: 
calculated contour plots of $m_z (x,y)$ in three layers with $\Delta \psi_c = 2 \pi/3$.
Details of the magnetization distribution are given in Fig. \ref{fig:crank2}.
\label{fig:crank1}
}
\end{figure}
\begin{figure*}
\includegraphics[width=2.0\columnwidth]{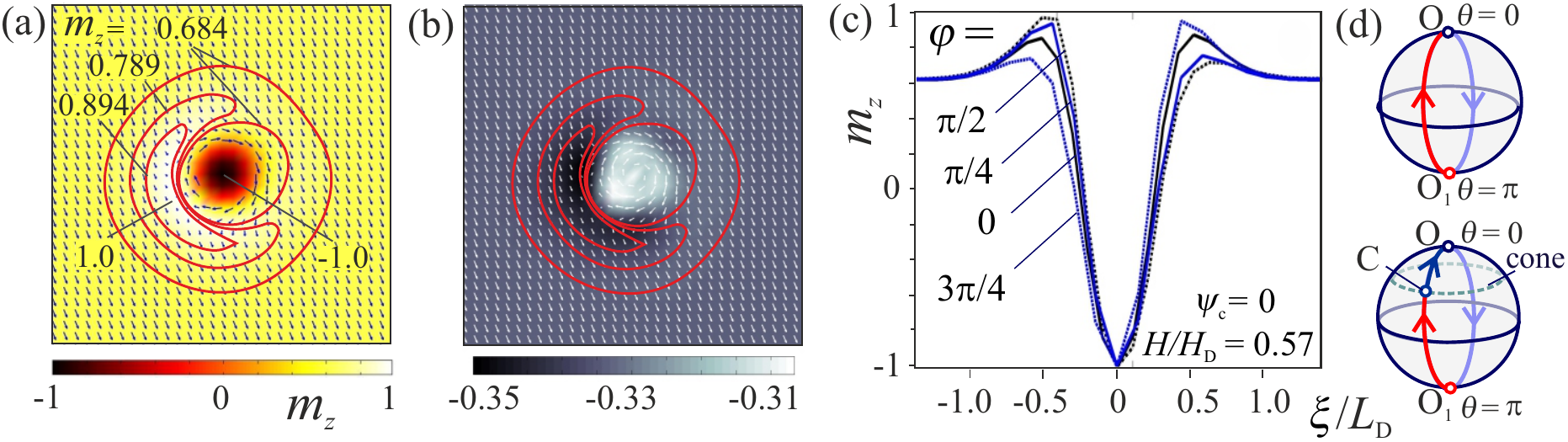}
\caption{ 
(color online) Numerical solutions for asymmetric skyrmions  (\ref{boundary2})
for $H = 0.57 H_D$: contour plots of $m_z (x,y)$ (a) and energy density $e (x,y)$
(b) in a $xy$ plane with a fixed value of $\psi_c$ (\ref{cone}); 
(c) magnetization profiles $\theta (r)$ for different values of $\varphi$ in the $xy$
layer with $\psi_c$ = 0 ($\xi$ is the spatial variable along lines with fixed
values of $\varphi$); (d) trajectories of the magnetization vector $\mathbf{m}$ 
for axisymmetric skyrmions (\ref{skyrmion0}) (top) and for asymmetric localized solitons 
(\ref{boundary2}) along the line $\varphi = \psi_c +\pi/2$ (bottom).
\label{fig:crank2}
}
\end{figure*}
\begin{figure*}
\includegraphics[width=2.0\columnwidth]{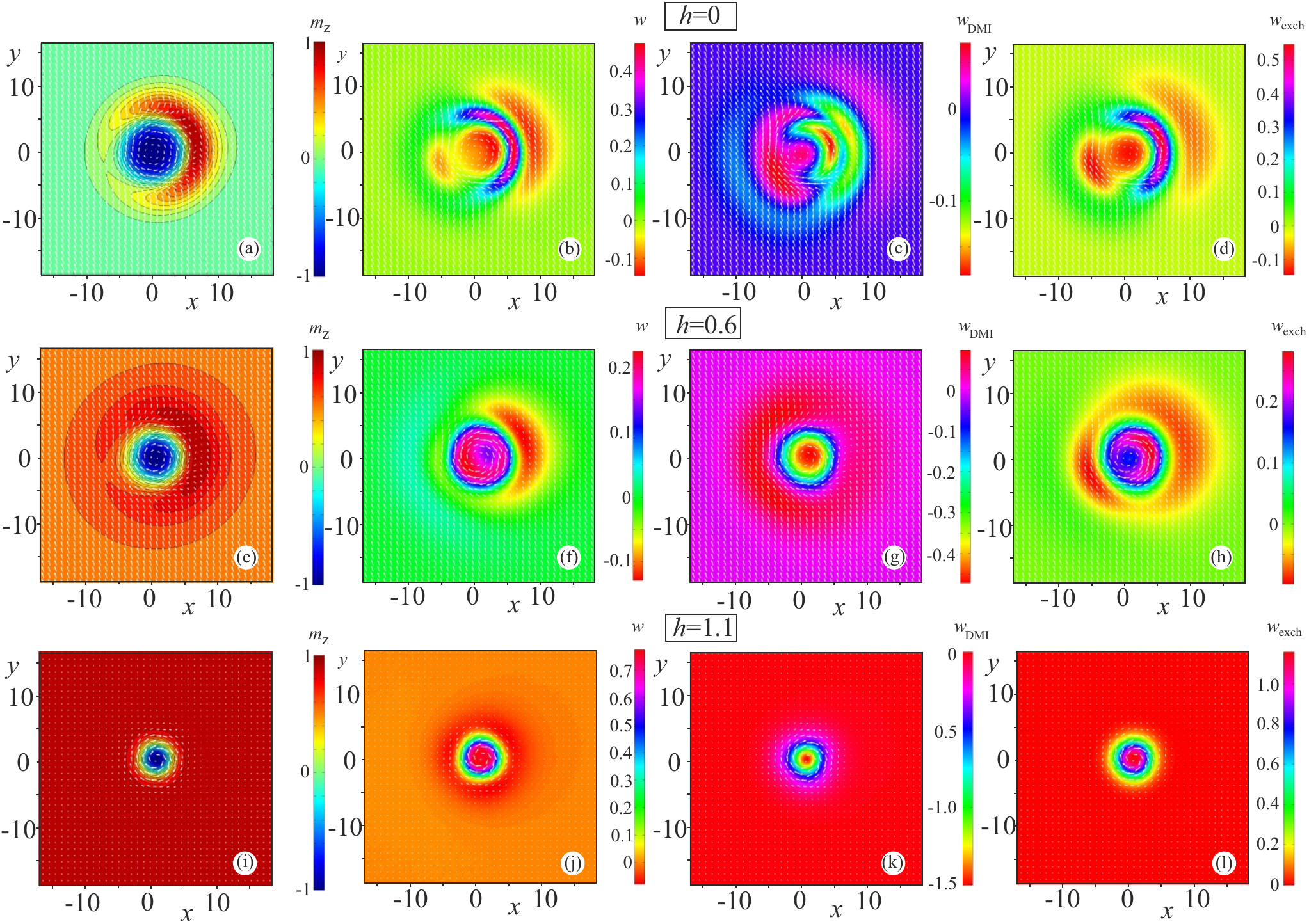}
\caption{
(color online) Numerical solutions for asymmetric skyrmions  (\ref{boundary2}) obtained within the continuum model (\ref{density})
for different values of the applied magnetic field. Contour plots of $m_z (x,y)$ (a), (e), (i), energy density $w$ (b), (f), (j), energy density of DM (c), (g), (k),  and exchange interaction  (d), (h), (l) are plotted in a $xy$ plane with a fixed value of $\psi_c$ (\ref{cone}).
\label{fig:crank33}
}
\end{figure*}

In our report we address this  compatibility problem
and derive regular solutions for asymmetric skyrmions
embedded into the cone phase.
We demonstrate that  unlike the  repulsive axisymmetric skyrmions
existing in the saturated states, chiral solitons in the cone phase
have an attractive interskyrmion potential and form biskyrmion 
and multiskyrmion states.

We consider the standard model for magnetic states
in cubic non-centrosymmetric ferromagnets \cite{Dz64,Bak80},
\begin{eqnarray}
w =A\,(\mathbf{grad}\,\mathbf{m})^2 + 
D\,\mathbf{m}\cdot \mathrm{rot}\,\mathbf{m}
- \mu_0 M \mathbf{m}\cdot \mathbf{H},  
\label{density}
\end{eqnarray}
where $\mathbf{m}= (\sin\theta\cos\psi;\sin\theta\sin\psi;\cos\theta)$
is the unity vector along the magnetization $\mathbf{M}$, 
$A$ is the exchange stiffness constant, 
$D$ is the Dzyaloshinskii-Moriya (DM) coupling energy,
and $\mathbf{H}$ is the applied magnetic field.

Chiral modulations along the applied field  with 
the period $L_D = 4 \pi A/|D|$ correspond to the global 
minimum of the functional (\ref{density}) below the critical 
field $ \mu_0 H_D = D^2/(2 A M)$.
The equilibrium parameters for the cone phase 
are expressed in analytical form \cite{Bak80}  as:
\begin{eqnarray}
 \theta_c = \arccos \left( H/H_D \right) , \quad \psi_c = 2\pi z/L_D,  \quad
\label{cone}
\end{eqnarray}
 where $z$ is the spatial variable along the applied field.

At $H = H_D$,  the cone phase transforms into
the saturated state with $\theta = 0$.
Within the saturated phase ($H > H_D$), isolated chiral skyrmions are
described by axisymmetric solutions of type 
\begin{eqnarray}
\theta = \theta (\rho), \quad \psi = \varphi + \pi/2, \quad
\label{skyrmion1}
\end{eqnarray}
that are homogeneous along the skyrmion axis $z$
where $\mathbf{r}$ = $(\rho \cos \varphi, \rho \sin \varphi, z)$ 
are cylindrical coordinates for the spatial variable \cite{JMMM94}.
The equilibrium solutions for $\theta (\rho)$ are derived from 
the Euler equation for the energy functional \cite{JMMM94}
 \begin{eqnarray}
\textstyle
 w_s (\theta)   =   A   \mathcal{J}_0  ( \theta )   
+ D  \mathcal{I}_{0} ( \theta ) -  \mu_0 M H \cos \theta, \quad \quad
\nonumber \\
\mathcal{J}_{0} ( \theta )   =  \theta_{\rho}^2  +\frac{1}{\rho^2}\sin^2 \theta, \
\mathcal{I}_{0} ( \theta )  = \theta_{\rho}+ \frac{1}{\rho} \sin \theta \cos \theta, \
\label{skyrmion0}
\end{eqnarray}
with the boundary conditions 
$\theta (0) = \pi$, $\theta (\infty) = 0$ (see Fig. \ref{fig:crank33} (i)-(l) for the distribution of the $m_z$-component of the magnetization and energy density distributions in a xy plane with a fixed value of $z$). 

Below the saturation field ($H < H_D$), the structure of two-dimensional 
skyrmions is imposed by the arrangement of the cone phase (\ref{cone}).
These solutions should be periodic with period $L_D$ along the $z-$ axis 
and are confined by the following  in-plane boundary conditions: 
\begin{eqnarray}
\theta_{\rho = 0} = \pi, \quad \theta_{\rho = \infty} =  \theta_c, \quad  
 \psi_{\rho = \infty} (z)  =  \psi_c (z).  \quad
\label{boundary2}
\end{eqnarray}
The solutions for $\theta (\rho, \varphi, z)$, $\psi (\rho, \varphi, z)$ 
are derived by minimization of the energy functional (\ref{density}):
\begin{eqnarray}
\!w \! = \! A \mathcal{J}\!( \theta, \psi) \!
+\! D \mathcal{I}\!( \theta, \psi)\!-\!\mu_0 M\!H\!\cos\!\theta,\ 
\label{energy2}
\end{eqnarray}
with the boundary conditions (\ref{boundary2}),
where the  exchange ($\mathcal{J}$) and Dzyaloshinskii-Moriya 
($\mathcal{I}$) energy functionals  are 

$\mathcal{J}  ( \theta, \psi)   =  \theta_{\rho}^2 + \theta_z^2 
+ \frac{1}{\rho^2}  \theta_{\varphi}^2 +
\sin^2 \theta \left(\psi_{\rho}^2+ \psi_z^2 +\frac{1}{\rho^2}  \psi_{\varphi}^2  \right), \quad $

$\mathcal{I}  ( \theta, \psi) =  \sin (\psi- \varphi)
 ( \theta_{\rho}+ \frac{1}{\rho} \sin \theta \cos \theta \: \psi_{\varphi}) 
+ \sin^2 \theta \: \psi_z \quad$
$\quad + \cos (\psi- \varphi)
 ( \frac{1}{\rho} \theta_{\varphi}+   \sin \theta \cos \theta \: \psi_{\rho})$.
%
%
%

 To investigate the solutions  for asymmetric skyrmions, we
 use  both the continuous and the discretized versions of Eq. (\ref{density}). 
In the continuous version, we use finite-difference approximation of derivatives in Eq. (\ref{density}) with different precision up to
eight points as neighbors on rectangular grids with adjustable grid spacings.
The size of the grid is up to 150x150x100.
In the discretized version, we consider classical spins of unit length on a three-dimensional cubic
lattice with the following energy functional: 
\begin {align}
w = &-\,\sum_{<i,j>} (\mathbf{S}_i \cdot \mathbf{S}_j ) -\sum_{i} \mathbf{h} \cdot \mathbf{S}_i 
 \nonumber\\
&- d \, \sum_{i}(\mathbf{S}_i \times \mathbf{S}_{i+\hat{x}} \cdot \hat{y} 
- \mathbf{S}_i \times \mathbf{S}_{i+\hat{y}} \cdot \hat{x})
\label{discrete}
\end{align}
where $\mathbf{h}=\mathbf{H}J/D^2$ and $d=J/D= 1/\tan (2π/p)$. $<i,j>$ denotes pairs of nearest-neighbor spins. 
The Dzyaloshinskii-Moriya constant $d$ defines the period of modulated structures $p$. Or vice versa, one choses the period of the modulations for the computing procedures and defines the corresponding value of $d$. 
In what follows, the Dzyaloshinskii-Moriya constant is set to 0.7265 which  corresponds to one-dimensional modulations with a period of 10 lattice spacings in zero field.  
The discrete  model (\ref{discrete}) is particularly useful when the continuum model becomes invalid for localized  solutions with sizes of few lattice constants \cite{Leonov15b,Leonov15}. The model also allows to operate with smaller arrays of spins as compared  with the continuum model. 

Numerical calculations for $H = 0.57 H_D$ (Figs. \ref{fig:crank1},
\ref{fig:crank2})  show the  main features of asymmetric
skyrmions.  The entire structure of a skyrmion within the cone phase
can be thought of as a stack of layers (Fig. \ref{fig:crank1})
rotating around the $z$ axis with period $L_D$.
These specific solitonic states are characterized by
three-dimensional chiral modulations: the cone modulations
along their axis and a double-twist rotation in the perpendicular
plane.

Figs. \ref{fig:crank2} (a-c) present the equilibrium structures of Eq. (\ref{discrete}) within
a layer with a fixed value of $z$.  The function $m_z (\rho, \varphi)$
consists of a strongly localized axisymmetric core separated from the
outer region with a fixed value of the magnetization $\mathbf{m}
(\theta_c, \psi_c)$ (\ref{cone}) by a broad strongly asymmetric
transitional region  we call the \textit{shell}.  

The contour lines of equal $m_z$ in Fig. \ref{fig:crank2} (a) and
magnetization profiles along the skyrmion diameter, $m_z (\xi)$  
in Fig. \ref{fig:crank2} (c), display details of the  shell.   
Note especially that  the
magnetization profile along $\varphi = \psi_c + \pi/2$ reaches the
value $\theta = 0$.  In Fig. \ref{fig:crank2} (a), this point is
enclosed by crescent-shaped contours.
The unit spheres in Fig. \ref{fig:crank2} (d) 
demonstrate the difference between the solution 
for axisymmetric skyrmions in the saturated phase
(\ref{skyrmion1}) and those within the cone phase (\ref{boundary2}).
The former are described by $O O_1$ lines connecting
the skyrmion center  with $\theta = \pi$ ($O_1$) and
the point $O$ corresponding to the system ``vacuum'', $\theta = 0$. 
Solutions for skyrmions within the cone phase
 are described by the lines connecting the skyrmion center
($O_1$) with the point  $C$ corresponding to the cone phase (\ref{cone}).
In Fig. \ref{fig:crank2} (d) we indicate the magnetization
trajectory along  $\varphi = \psi_c + \pi/2$ direction.
Fig. \ref{fig:crank33} (the first column) shows the contour plots of $m_z$ components refined within the continuous model (\ref{density}) for different values of the applied magnetic field.
The second column shows the energy density $w$ of Eq. (\ref{density}) while the third and the fourth column show DM and exchange energy densities, correspondingly.

\begin{figure}
\includegraphics[width=0.99\columnwidth]{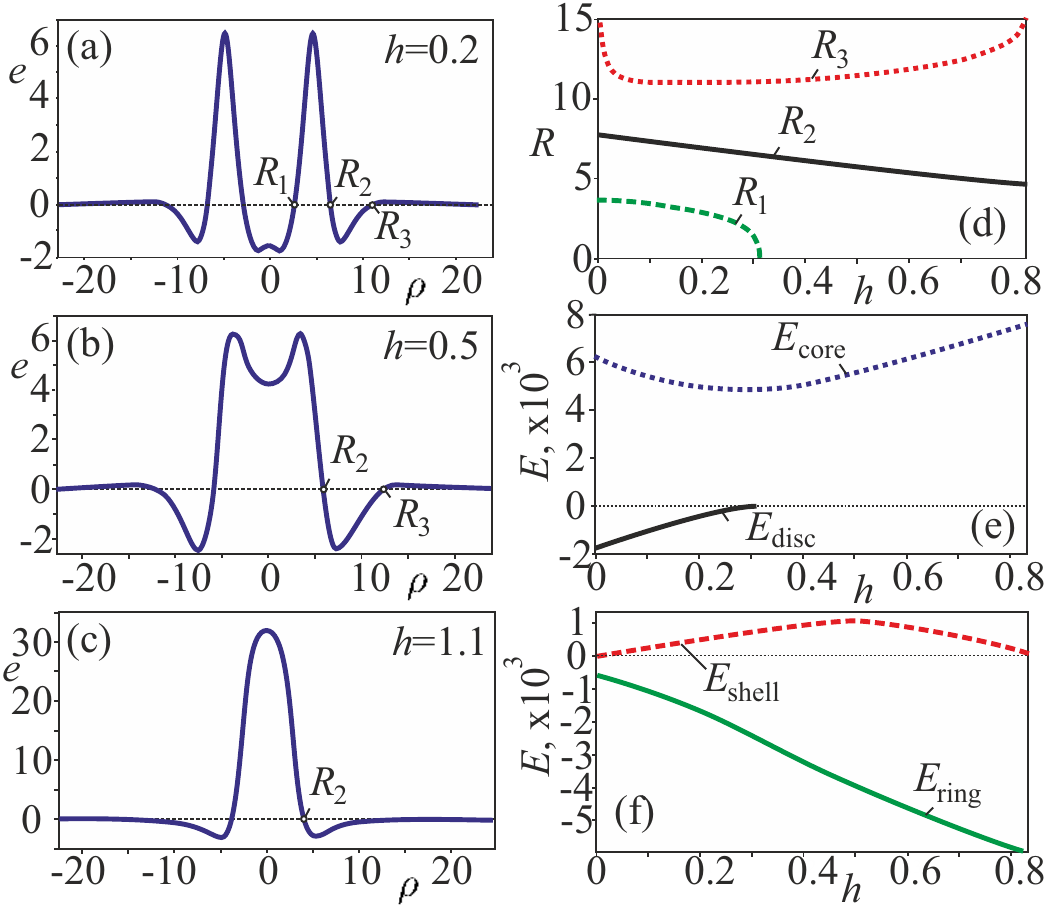}
\caption{
(color online). Variation of the energy density $e (\rho)$ along the skyrmion diameter, 
 for different values of the applied magnetic field (a), (b), (c).
(d) dependence of the characteristic radii $R_1$, $R_2$, $R_3$ of the energy density $e (\rho)$ from (a)-(c) on the field.
The total energies $E$ accumulated within the different parts of the asymmetric skyrmions in dependence on the field (e), (f).
The Energy of the \textit{disc} $E_{\rm{disc}}$ was integrated over the interval $[0,R_1]$, the energy of the \textit{core} $E_{\rm{core}}$ - over $[0,R_2]$, the energy of the \textit{ring} $E_{\rm{ring}}$ - in the interval $[R_2,R_3]$, and the energy of the \textit{shell} $E_{\rm{shell}}$ - for $\rho>R_3$.
\label{fig:crank4}
}
\end{figure}
\begin{figure}
\includegraphics[width=0.93\columnwidth]{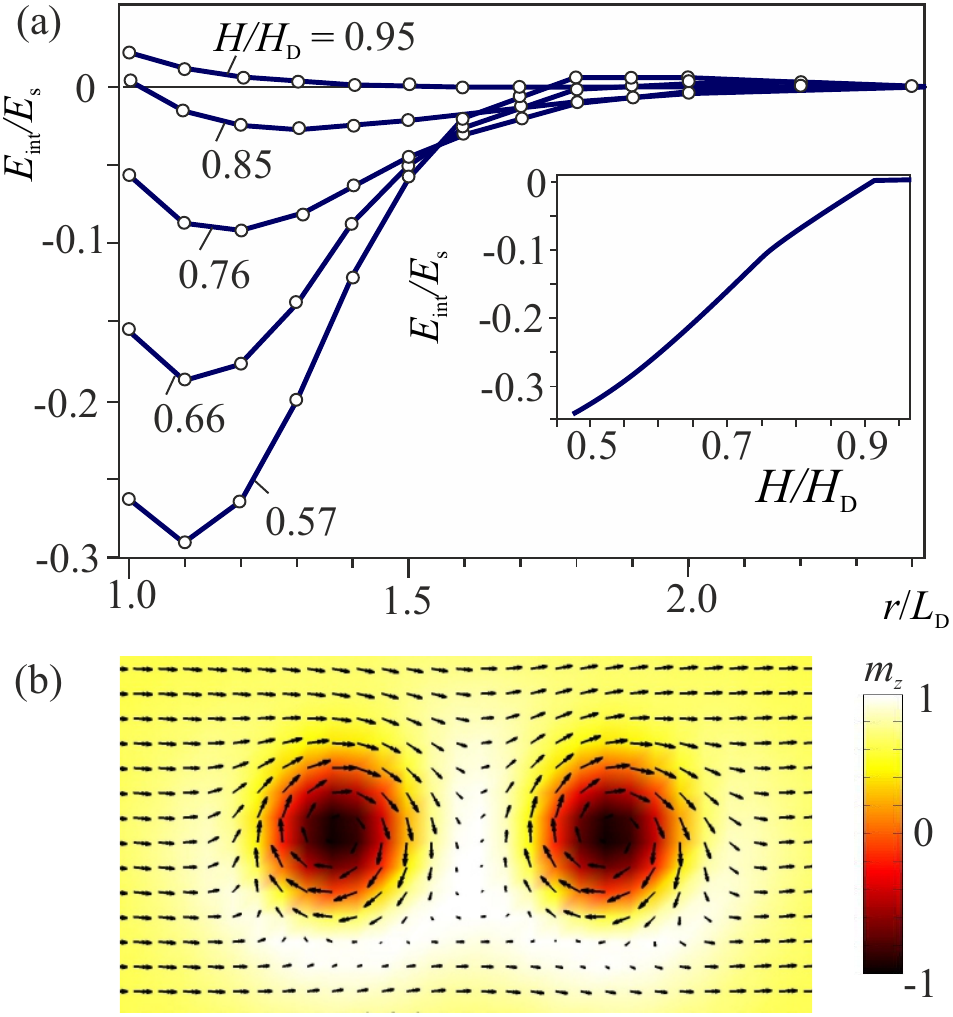}
\caption{
(color online). 
Reduced energy for the interaction between two asymmetric skyrmions,
$E_{int}/E_{s}$,  as a function of the distance between the skyrmion
centers $r$ calculated for different values of the applied field (a). 
Inset shows minimal values of $E_{int}$ as a function of the applied field.
Contour plot of $m_z (x,y)$ for bound asymmetric skyrmions
 at $H/H_D$ = 0.57 (b).
\label{fig:crank5}
}
\end{figure}
Fig. \ref{fig:crank4} shows the radial skyrmion energy density $e(\rho) = (2 \pi L_D)^{-1} 
\int_0^{L_D} dz \int_0^{2\pi} d \varphi w_s (\theta, \psi)$ for different values of the field as calculated with respect to the energy density of the conical phase. 
The positive exponentially decaying asymptotics of $e(\rho)$ (Fig. \ref{fig:crank4} (a), (b)) imply the
\textit{attractive} interaction between the skyrmions in  the cone phase, whereas the axisymmetric skyrmions in the
saturated state of chiral magnets (Fig. \ref{fig:crank4} (c)) have a repulsive interskyrmion potential \cite{JETPL95,JPCS11,Komineas15}.
$e(\rho)$ has three characteristic radii $R_1$, $R_2$, and $R_3$, dependence of which on the field is plotted in Fig. {\ref{fig:crank4}} (d).
The \textit{shell} mentioned above is the part of the asymmetric skyrmion with $\rho>R_3$. Note, that the \textit{shell} disappears in zero field as well as for $h>0.8$. 
The characteristic radius $R_2$ specifies the size of the skyrmionic \textit{core} ($0<\rho<R_2$). 
We also call the part of the skyrmion with $R_2<\rho<R_3$  the \textit{ring}. 
In Ref. \cite{JPCS11} it was pointed out that this ring with the negative energy density due to the DM interaction guarantees the stability of the chiral skyrmion. 
Note that with decreasing magnetic field a \textit{disc} ($\rho<R_1$) with the negative energy density develops. 
The total energies $E$ accumulated within the different parts of the asymmetric skyrmions are shown in dependence on the field in Fig. \ref{fig:crank4} (e), (f).

In Fig. \ref{fig:crank5} (a) the reduced interaction energy 
between two asymmetric skyrmions, $E_{int}/E_{s}$  
is plotted as a function of their separation distance for
different values of the applied magnetic field
($E_s = \int_0^{\infty} e (\rho) \rho d\rho$ is the total
equilibrium energy of an isolated asymmetric skyrmion).
The Lennard-Jones type potential profiles $E_{int}/E_{s} (r/L_D)$  
show that the attractive interskyrmion coupling
is characterized by a low potential barrier and a rather
deep potential well establishing the equilibrium separation
of skyrmions in the bound biskyrmion state 
(Fig. \ref{fig:crank5} (b)).

In  conclusion, regular solutions for
isolated chiral skyrmions in the cone
phase of cubic helimagnets have been derived
by numerically solving   the corresponding
micromagnetic equations (\ref{boundary2})  and  (\ref{energy2}).
These novel solitonic states are characterized by
three-dimensional chiral modulations and an attractive
interskyrmion potential.
Similar skyrmionic states can  arise in
the cone phases admissible in uniaxial chiral 
ferromagnets with $C_n$ and $D_n$ symmetry 
\cite{Dz64,JETP89}.
Axisymmetric skyrmions exist in the saturated phase
of chiral ferromagnets as ensembles of repulsive isolated
particles \cite{JMMM94,JPCS11,Romming13}.
Our findings show that below the transition field into
the cone phase, the axisymmetric skyrmions transform
into asymmetric attractive solitons and may form biskyrmion
and multiskyrmion states (clusters). 

To date, no direct observations of isolated skyrmions or skyrmion
clusters have been reported in the cone phases of chiral
helimagnets. However, a few isolated observations  such as
a decomposition of a skyrmion lattice into cluster-like patterns
\cite{Yu10} and the formation of skyrmionic droplets in MnSi plates
\cite{Grigoriev14},  are in accord with our theoretical results
and indicate possible directions for the investigation of this
phenomenon.

The authors are grateful to K. Inoue, J. Kishine,
and G. Tatara for useful discussions.
A.O.L acknowledges financial support by the FOM
grant 11PR2928.  A.N.B acknowledges support  
by the Deutsche Forschungsgemeinschaft 
via Grant No. BO 4160/1-1.


\begin{thebibliography} {99}

\bibitem{Dz64}  
Dzyaloshinskii I E 1964 Sov. Phys. JETP \textbf{19} 960

\bibitem{Bak80}
Bak P and Jensen M H 1980 J. Phys.C: Solid State Phys. \textbf{13} L881

\bibitem{JETP89}
Bogdanov A N  and Yablonsky A D 1989 Sov. Phys. JETP \textbf{68} 101

\bibitem{JMMM94}
Bogdanov A and Hubert A 1994 J. Magn. Magn. Mater. \textbf{138} 255

\bibitem{Ishikawa76} 
Ishikawa Y, Tajima K, Bloch D and Roth M 1976 Solid State
Commun. \textbf{19} 525

\bibitem{Beille81} 
Beille J, Voiron J, Roth M and Zhang Z Y 1981 
J. Phys. F: Met. Phys. \textbf{11} 2153

\bibitem{Lebech89} 
Lebech B, Bernhard J and Freltoft T 1989 
J. Phys.: Condens. Matter \textbf{1} 6105

\bibitem{Togawa12}
Togawa Y, Koyama T,  Takayanagi K,  Mori S,  Kousaka Y, Akimitsu J, 
Nishihara S,  Inoue K,  Ovchinnikov A S and  Kishine J I, 2012
Phys. Rev. Lett. {\textbf{108}} 107202 

\bibitem{Porter15}
Porter N A \textit{et al} 2015 Phys. Rev. B {\textbf{92}} 144402 


\bibitem{Romming13} 
Romming N, Hanneken C,  Menzel M, Bickel J E, 
Wolter B, von Bergmann K, Kubetzka A, and  Wiesendanger R 2013
  Science {\textbf{341}} 636 
	
\bibitem{Romming15} 
Romming N, Kubetzka A, Hanneken C,  von Bergmann K,  Wiesendanger R 2015
  Phys. Rev. Lett.  {\textbf{114}} 177203 ;
	Marrows C H 2015 Physics {\textbf{8}} 40

\bibitem{Leonov15} A. O. Leonov, T. L. Monchesky, N. Romming, A. Kubetzka, A. N. Bogdanov, R. Wiesendanger arXiv: 1508.02155.


\bibitem{JETPL95}
Bogdanov A 1995 JETP Lett. \textbf{62} 247 

\bibitem{JPCS11}
 R\"{o}{\ss}ler U K,  Leonov A A,  Bogdanov A N
2011 J. Phys.: Conf. Ser. \textbf{303} 012105

\bibitem{Leonov15b} Leonov A O, Mostovoy M 2015 Nat. Commun. \textbf{6} 8275


\bibitem{Komineas15}
Komineas S and Papanicolaou N 2015
 Phys. Rev. B \textbf{92} 064412 

\bibitem{Yu10}
Yu X Z, Onose Y, Kanazawa N, Park J H,  Han J H,
 Matsui Y,  Nagaosa N and  Tokura Y 2010
Nature (London) \textbf{465} 901 

\bibitem{Grigoriev14}
Grigoriev S V,   Potapova N M,   Moskvin E V, Dyadkin V A,  Dewhurst Ch and  Maleyev S V 2014
JETP Lett. \textbf{100} 216 






\end{thebibliography}
\end{document}